\documentstyle[preprint,epsfig,aps]{revtex}
\begin{document}
\draft
\preprint{\vbox{\hbox{SNUTP 94-74}\hbox{KWUTP-94/1}}}

\title{
Statistical Mechanics of Non-Abelian Chern-Simons Particles}

\author{Taejin Lee\cite{tlee}}

\address{
Department of Physics, Kangwon National University, Chuncheon 200-701, KOREA }

\maketitle
\begin{abstract}

We discuss the statistical mechanics of a two-dimensional gas of non-Abelian
Chern-Simons particles which obey the non-Abelian braid statistics.
The second virial coefficient is evaluated in the framework of the non-Abelian
Chern-Simons quantum mechanics.
\end{abstract}


\narrowtext


One of the novel features of anyons \cite{anyon} is that their thermodynamic
character as well as their spin and statistics interpolates between that of
bosons to that of fermions as the statistical parameter varies.
This feature has been clearly exhibited in the works of Arovas, Schrieffer,
Wilczek and Zee \cite{arovas} where they studied the two-dimensional gas of
free
anyons in the low-density regime by taking the virial expansion. In
particular, the second virial coefficient was evaluated in their study and was
found to have periodic, non-analytic behavior as a function of the statistical
parameter.  Comtet, Georgelin and Ouvry \cite{comtet} confirmed this result by
making use of a harmonic potential as means of a regulator and Blum, Hagen and
Ramaswamy \cite{blum} discussed further the effects of spin on the second
virial
coefficient.

In this letter, we will discuss the statistical mechanics of a two dimensional
gas of $SU(2)$ non-Abelian Chern-Simons (NACS) particles in the framework of
the non-Abelian Chern-Simons quantum mechanics which has been developed
recently
in refs.\cite{leeoh1,leeoh2,bjp}. The NACS particles are point-like sources
which interact with each others through a topological non-Abelian
Aharonov-Bohm
interaction. Carrying both non-Abelian charges and non-Abelian magnetic fluxes,
they acquire fractional spins and obey braid statistics. Since the braid
(non-Abelian) statistics  is much more generalized than the fractional
statistics of anyons, it is interesting to explore their statistical mechanical
properties and to compare them with those of anyons. The NACS particle and
the non-Abelian Aharonov-Bohm effect have been discussed in connection
with various physical phenomena such as the scattering \cite{vort} of
non-Abelian vortices which form in some spontaneously broken gauge
theories, the fractional quantum Hall effect \cite{fqh}, and the (2+1)
dimensional gravity \cite{3gr}. Thus, exploring the statistical mechanics
of the NACS particles is important in many respects.

The dynamics of the $N$-body system of  free NACS particles are governed by the
Hamiltonian \cite{leeoh1,leeoh2,ver}
\begin{eqnarray}
{H}_N&=&-\sum_{\alpha=1}^{N} {1\over \mu_\alpha}\left(\nabla_{\bar
z_\alpha}\nabla_{z_\alpha}  +\nabla_{z_\alpha}\nabla_{\bar
z_\alpha}\right) \nonumber \\
\nabla_{z_\alpha}&=&{\partial\over \partial z_\alpha}  +{1\over 2\pi
\kappa} \sum_{\beta\not=\alpha} \hat Q^a_\alpha \hat Q^a_\beta {1\over
z_\alpha -z_\beta}\label{ham} \\
\nabla_{\bar z_\alpha}&=&{\partial\over \partial \bar z_\alpha} \nonumber
\end{eqnarray}
where the particles are labeled by $\alpha = 1, \dots, N$ and their spatial
coordinates are denoted by $(q^1_\alpha, q^2_\alpha)=(z_\alpha+\bar z_\alpha,
-i(z_\alpha-\bar z_\alpha))/2$. Here $\kappa$ is a parameter of the theory such
that $4\pi\kappa= {\rm integer}$ and $\hat Q^a$'s are the
isovector operators which can be represented by some generators $T^a_l$ in a
representation of isospin $l$. One can construct a classical model
\cite{leeoh1,leeoh2,bjp,bal90} where the NACS particles are
described as isospin particles \cite{wong,bal78} of which isospin charges  are
minimally coupled with the  non-Abelian Chern-Simons gauge fields \cite{deser}.
Solving the Gauss' constraint explicitly and integrating out the gauge fields
by
choosing the holomorphic gauge \cite{leeoh1,leeoh2} in the framework of the
coherent state quantization \cite{coh1,coh2}, one obtains the Hamiltonian
$H_N$,
Eq.(\ref{ham}).

We may remove the interaction terms in $H_N$ by a similarity transformation
\begin{eqnarray}
{H}_N&\longrightarrow & UH_N U^{-1}= H^{\rm free}_N = -\sum^N_\alpha
\frac{2}{\mu_\alpha} \partial_{\bar z_\alpha}\partial_{z_\alpha}\nonumber\\
\Psi_H &\longrightarrow & U \Psi_H = \Psi_A \label{simil}
\end{eqnarray}
where $U(z_1,\dots,z_N)$ satisfies the Knizhnik-Zamolodchikov (KZ)
equation \cite{kz}
\begin{equation}
\left({\partial\over \partial z_\alpha}  - {1\over 2\pi
\kappa} \sum_{\beta\not=\alpha} \hat Q^a_\alpha \hat Q^a_\beta {1\over
z_\alpha -z_\beta}\right) U(z_1,\dots,z_N) =0\label{kzeq}
\end{equation}
and $\Psi_H(z_1,\dots,z_N)$ denotes the wave function of the $N$-body system
of the NACS particles in the holomorphic gauge.
Comparing Eq.(\ref{kzeq}) with the KZ equation which is
satisfied by the Green's functions in the conformal field theory, we
find that ($4\pi\kappa - 2$) corresponds to the level of the underlying $SU(2)$
current algebra. Note that $\Psi_A$ obeys the braid statistics
due to the transformation function $U(z_1,\dots,z_N)$ \cite{leeoh1,leeoh2}
while $\Psi_H$ satisfies ordinary statistics.
In analogy with the Abelian Chern-Simons particle theory
$\Psi_A$ may be called the NACS particle wave function in the anyon gauge.
The transformation function also defines the inner product in
the holomorphic gauge
\begin{equation}
<\Psi_1 |\Psi_2> = \int d^{2N}\zeta \Psi_1(\zeta)^\dagger U^\dagger
(\zeta) U(\zeta) \Psi_2 (\zeta)\label{inner}
\end{equation}
where $\zeta = (z_1,\dots,z_N)$. This inner product renders the Hamiltonian
in the holomorphic gauge $H_N$ Eq.(\ref{ham}) Hermitian, which does not look
manifestly Hermitian.

Defining a matrix (operator) valued 1-from
\begin{equation}
\omega = \sum_{\alpha<\beta}\frac{1}{2\pi\kappa} \Omega_{\alpha\beta}\, d \,
\ln (\zeta_\alpha -\zeta_\beta)
\end{equation}
where $\Omega_{\alpha\beta} = \hat Q^a_\alpha \hat Q^a_\beta$,
we find that the transformation function $U$ corresponds to the monodromy of
$\omega$ \cite{mono}
\begin{equation}
U(z_1,\dots,z_N) = I + \int_\Gamma \omega + \int_\Gamma \omega\omega + \dots
\end{equation}
where $\Gamma$ is a contour in the $N$-dimensional complex space with one end
point-fixed and the other being $(z_1,\dots,z_N)$. The integrability condition
\begin{equation}
\left[ \nabla_{z_\alpha}, \nabla_{z_\beta}\right] = 0
\end{equation}
for the transformation function $U$ to exist leads to
\begin{equation}
d\omega + \omega \wedge \omega = 0.
\end{equation}
This condition is fulfilled, since $\Omega_{\alpha\beta}$ satisfy
the infinitesimal pure braid relations \cite{mono}
which are relevant to the classical Yang-Baxter equation.

Having defined the non-Abelian Chern-Simons quantum mechanics, we turn to the
statistical mechanics of the NACS particles. The grand partition function $\Xi$
is defined as usual in terms of the $N$-body Hamiltonian $H_N$ and
the fugacity $\nu$ by
\begin{equation}
\Xi = \sum_{N=0}^\infty \nu^N\, {\rm Tr}\, e^{-\beta H_N}\label{par1}
\end{equation}
where $\beta=1/kT$. In the low-density regime, a cluster expansion can be
applied to $\Xi$
\begin{equation}
\Xi = \exp \left( V\sum_{n=1}^\infty b_n \nu^n\right)\label{par2}
\end{equation}
where $V$ is the volume (the area $A$ for the two-dimensional gas) and $b_n$ is
the $n$-th cluster integral. Comparing the two expressions for $\Xi$,
Eqs.(\ref{par1}) and (\ref{par2}), we have
\begin{equation}
b_1 = \frac{1}{V} Z_1,\quad b_2 = \frac{1}{V}\left(Z_2 - Z_1^2/2\right)
\end{equation}
and
\begin{equation}
Z_N = \frac{1}{N!}\int d^{2N}\zeta \,
\sum_{\{m_i\}} \sum_P <(m_1,z_1)\dots (m_N,z_N)|e^{-\beta H_N}
|P(m_1,z_1) \dots P(m_N,z_N)>\label{part}
\end{equation}
where $m_i$, $i=1,\dots,N$ denotes the isospin quantum number of the
$i$-th particle.
We note that $P$ denotes the operation of permutation and the
nontrivial inner product Eq.(\ref{inner}) is to be appropriately
taken into account in Eq.(\ref{part}).

Since the NACS particles are described in the regular gauges, for instance,
the Coulomb, axial and holomorphic gauges, as bosons interacting
with each other through the topological terms induced by the
Chern-Simons gauge fields, the expression of the $N$-particle partition
function Eq.(\ref{part}) holds generally in cases of the regular gauges.
However, in the anyon gauge where the non-Abelian braid statistics of the
NACS becomes manifest, this expression for $Z_N$ is not valid and
some modification is necessary. We may rewrite Eq.(\ref{part}) as
\begin{eqnarray}
Z_N &=& \sum_{\{m_i\}} \int d^{2N}\zeta \, \Psi^*_{\{m_i\}}(\zeta)
U^\dagger(\zeta)U(\zeta) e^{-\beta H_N} {\bf P}
\Psi_{\{m_i\}}(\zeta),\label{ZN}\\
{\bf P} &=& \frac{1}{N!}\sum^{N-1}_{i=0} S_i,\nonumber
\end{eqnarray}
in terms of the exchange operators; $S_0 = I$ and $S_i$, $i=1,\dots,N-1$
\begin{eqnarray}
S_i &\Psi&_{m_1,\dots,m_i,m_{i+1},\dots,m_N}
(z_1,\dots,z_i,z_{i+1},\dots,z_N) \nonumber\\
& &\quad =\Psi_{m_1,\dots,m_{i+1},m_{i},\dots,m_N}
(z_1,\dots,z_{i+1},z_{i},\dots,z_N)\label{exch}
\end{eqnarray}
where $\Psi_{\{m_i\}}(\zeta)$ denotes
$\Psi_{m_1,\dots,m_N} (z_1,\dots,z_N)= <(m_1,z_1)\dots (m_N,z_N)|\Psi>$.
In the anyon gauge we may obtain a similar expression for $Z_N$. But
the exchange operator is no longer represented by the simple
permutation $P$ and a nontrivial exchange factor, which satisfies the
Yang-Baxter equation, must be introduced \cite{leeoh2}
\begin{equation}
S^A_i \Psi_{A} (z_1,\dots,z_i,z_{i+1},\dots,z_N)
={\cal R}(z_i, z_{i+1}) \Psi_{A} (z_1,\dots,z_{i+1},z_{i},\dots,z_N)
\end{equation}
where $\Psi_A$ denotes the wave function in the anyon gauge.
Thus, defining ${\bf P}^A = 1/N! \sum^{N-1}_{i=0} S^A_i$,
$S^A_0 = I$ we have
\begin{eqnarray}
Z_N &=& \sum_{\{m_i\}} \int d^{2N} \zeta \, \Psi^{A*}_{\{m_i\}}(\zeta)
e^{-\beta H^A_N} {\bf P}^A \Psi^A_{\{m_i\}}(\zeta)\label{ZNA}\\
&=& \int d^{2N} \zeta \,{\rm tr}\,<z_1,\dots,z_N|e^{-\beta H^A_N}
{\bf P}^A |z_1,\dots,z_N>\nonumber
\end{eqnarray}
where $H^A_N$ is the $N$-body Hamiltonian in the anyon gauge.

The virial expansion, i.e., the expansion of the equation of
state in powers of the density $\rho$ is given as
\begin{equation}
P= \rho kT\left( 1+ B_2(T) \rho+ B_3(T) \rho^2 +\dots \right)
\end{equation}
where $B_n(T)$ is the $n$-th virial coefficient.
The second virial coefficient $B_2(T)$ which is the main subject of this letter
is written as
\begin{equation}
B_2(T) = -\frac{b_2}{b_1^2}
= A\left(\frac{1}{2}-\frac{Z_2}{Z_1^2}\right). \label{btwo}
\end{equation}
Thus evaluation of $B_2(T)$ only involves the one-particle and two-particle
particle partition functions. If we assume that all the NACS particles belong
to
the same isospin multiplet $\{m=-l, \dots, l; |l,m>\}$ and have the same mass
$2\mu$,
\begin{equation}
Z_1 = {\rm Tr}\, e^{-\beta H_1} = (2l+1) A \lambda^{-2}_T \label{zone}
\end{equation}
where $\lambda_T = \sqrt{2\pi\hbar^2/\mu kT}$ is the thermal wavelength.
Since the two-body Hamiltonian can be expressed as
\begin{eqnarray}
H_2 &=& H_{\rm cm} + H_{\rm rel}\nonumber\\
&=&-{1\over 2\mu} \partial_Z \partial_{\bar Z}
-\frac{1}{\mu}(\nabla_z\nabla_{\bar z} +\nabla_{\bar z}\nabla_z),\label{two}\\
\nabla_z &=& \partial_z +\frac{\Omega}{z},\quad \nabla_{\bar z} =
\partial_{\bar z}\nonumber
\end{eqnarray}
where $Z = (z_1+z_2)/2$, $z = z_1 -z_2$ are the center of mass and the relative
coordinates respectively,
the two-particle partition function are factorized into
the contribution of the center of mass coordinates and that of the relative
coordinates as usual
\begin{mathletters}
\label{hamrel:all}
\begin{eqnarray}
Z_2 &=& {\rm Tr}\, e^{-\beta H_2} = 2A  \lambda^{-2}_T Z_2^\prime,
\label{hamrel:a} \\
Z_2^\prime &=& {\rm Tr}_{\rm rel}\, e^{-\beta H_{\rm rel}}\nonumber\\
&=& \frac{1}{2}\int d^2 z \sum_{m_1,m_2}\Bigl[
<m_1,m_2;z|e^{-\beta H_{\rm rel}}|m_1,m_2;z>\nonumber\\
& &\qquad \qquad+ <m_1,m_2;z|e^{-\beta H_{\rm rel}}|m_2,m_1;-z>
\Bigr].\label{hamrel:b}
\end{eqnarray}
\end{mathletters}
In Eq.(\ref{two}) $\Omega$ is a block-diagonal matrix given by
\begin{eqnarray}
\Omega &=& \hat Q^a_1\hat Q^a_2 / (2\pi\kappa)=\frac{1}{4\pi\kappa}
\left((\hat Q_1+\hat Q_2)^2-(\hat Q_1)^2-(\hat
Q_2)^2\right)\label{omega}\nonumber\\
&=& \sum_{j=0}^{2l}\frac{1}{4\pi\kappa}
\left(j(j+1)-2l(l+1)\right)\otimes{I}_j\\
&=& \sum_{j=0}^{2l} \omega_j\otimes{I}_j.\nonumber
\end{eqnarray}

As observed in the discussion on the scattering of the NACS particles
\cite{leeoh2}, it is convenient to take a similarity transformation given by
\begin{eqnarray}
H_{\rm rel} &\longrightarrow &
H_{\rm rel}^\prime = G^{-1} H_{\rm
rel} G,\nonumber\\
\Psi(z,\bar z) &\longrightarrow &
\Psi^\prime(z,\bar z) =G^{-1} \Psi(z,\bar z)\label{sim}
\end{eqnarray}
where
$G(z,\bar z)  = \exp\left(-\frac{\Omega}{2}\ln(z\bar z)\right)$.
This similarity transformation renders the inner product trivial and the
Hamiltonian $H_{\rm rel}^\prime$ becomes manifestly Hermitian
\begin{eqnarray}
H_{\rm rel}^\prime &=& -\frac{1}{\mu}(\nabla_z^\prime\nabla_{\bar z}^\prime
+ \nabla_{\bar z}^\prime\nabla_z^\prime),\label{trham}\\
\nabla_z^\prime &=& \partial_z + \frac{\Omega}{2}\frac{1}{z},\quad
\nabla_{\bar z}^\prime=\partial_{\bar z} -\frac{\Omega}{2}\frac{1}{\bar z}.
\nonumber
\end{eqnarray}
The partition function is invariant under this similarity transformation, i.e.,
$Z^\prime_2 = {\rm Tr}_{\rm rel}\, e^{-\beta H_{\rm rel}^\prime}$.
Rewriting the Hamiltonian $H_{\rm rel}^\prime$ in the polar coordinates
and projecting it onto the subspace of total isospin $j$, we see that
it corresponds to the Hamiltonian in the Coulomb gauge
for anyons of which statistical parameter is
given by $\alpha_s= \omega_j$
\begin{equation}
H_j^\prime = -\frac{1}{2\mu}\left[\frac{\partial^2}{\partial r^2}+
\frac{1}{r}\frac{\partial}{\partial r}+\frac{1}{r^2}\left(\frac{\partial}
{\partial \theta}+i\omega_j\right)^2\right].\label{hampr}
\end{equation}
Then it follows from symmetry of the Clebsch-Gordan
coefficients in the case of $SU(2)$, the two-particle partition
function may be written as
\begin{equation}
Z^\prime_2 = \frac{1}{2} \int d^2 z \sum^{2l}_{j=0}(2j+1)\left[
<z| e^{-\beta H^\prime_j} |z>+ (-1)^j <z| e^{-\beta H^\prime_j} |-z>\right].
\end{equation}

Introducing a harmonic potential $\frac{\mu}{2} \epsilon^2 r^2$ to the
Hamiltonian for the purpose of regularization, we find that the spectrum for
the
relative Hamiltonian $H_{j}^\prime$ is discrete \cite{anyon} and
can be classified into two classes: Type I with energy
$E^I_n = \epsilon (2n+1+\omega_j -[\omega_j])$, and degenercy of $n+1$ and
type II with energy $E^{II}_n = \epsilon (2n+1-\omega_j +[\omega_j])$, and
degenercy of $n$ where $n$ is a non-negative integer and $[\omega_j]$ is the
integer such that $0\le \delta_j= \omega_j-[\omega_j] < 1$.
In order to take an appropriate regularization \cite{arovas,comtet}
we define the second virial coefficient as
\begin{equation}
B_2(\kappa, l, T) - \Delta B_2(l,T) = -\frac{2\lambda_T^2}{(2l+1)^2}
\left[Z^\prime_2 (\kappa, l, T) - \Delta Z^\prime_2 (l,T)\right],
\end{equation}
where $\Delta B_2(l,T)$ is given in terms of the well-known virial coefficients
of Bose and Fermi systems ($B^B,\, B^F$) as
\begin{eqnarray}
\Delta B_2(l,T) & = & \frac{1}{(2l+1)^2} \sum^{2l}_{j=0} (2j+1)
\left[\frac{1+(-1)^j}{2}B^B(T)+ \frac{1-(-1)^j}{2}B^F(T) \right]\\
&= & -\frac{1}{4} \lambda^2_T \frac{1}{(2l+1)^2}
\sum^{2l}_{j=0} (2j+1)(-1)^j. \nonumber
\end{eqnarray}
The regularized partition function may be written by
\begin{eqnarray}
Z^\prime_2 (\kappa, l, T) - \Delta Z^\prime_2 (l,T) &=&
\sum_{j=0}^{2l}(2j+1)\lim_{\epsilon\rightarrow
0}\Biggl[\frac{1+(-1)^j}{2}\left(Z_\epsilon(\delta_j) -Z_\epsilon(0)\right)
\nonumber\\ & &\qquad\qquad +\frac{1-(-1)^j}{2}\left(Z_\epsilon(\delta_j)
-Z_\epsilon(1)\right)\Biggr],\\
Z_\epsilon(\delta_j)&=& \sum_{n=0}^{\infty}\left[(n+1) e^{-\beta\epsilon(2n+1
+\delta_j)}+n e^{-\beta\epsilon(2n+1-\delta_j)}\right] \nonumber\\
&=&\frac{1}{2} \frac{\cosh\left(\beta\epsilon(\delta_j-1)\right)}{\sinh^2
\beta\epsilon}.\nonumber
\end{eqnarray}
Then, by some algebra we get
\begin{equation}
B_2(\kappa, l, T) = -\frac{1}{4}\lambda^2_T+\frac{1}{(2l+1)^2}
\sum^{2l}_{j=0}(2j+1) \left[\delta_j-\frac{1}{2}\delta^2_j \right]
\lambda^2_T. \label{virial2}
\end{equation}

Since the NACS particle is a generalization of the anyon, one may expect that
the thermodynamic properties of the NACS particle are similar to those of the
anyon. However, this is not the case. Noting that both carry anomalous spins,
which are given by $s=l(l+1)/2\pi\kappa$  for the NACS particle and
$s=\alpha_s/2$ for the anyon, we may compare the behavior of the
virial coefficient of the NACS particle as a function of the spin with
that of anyon. We find that the periodicity in $s$, which is the case for the
anyon, does no longer hold and its functional behavior radically differs  from
that of the anyon even in the large-$\kappa$ limit as depicted in Fig.1.

\acknowledgments

This work was supported in part by the Basic Science Research
Institute Program, Ministry of Education, Korea (BSRI-94-2401),
by non-directed research fund, Korea Research Foundation 1994
and by KOSEF through the Center for Theoretical Physics at
Seoul National University.

\begin{figure}
\centerline{\epsfig{file=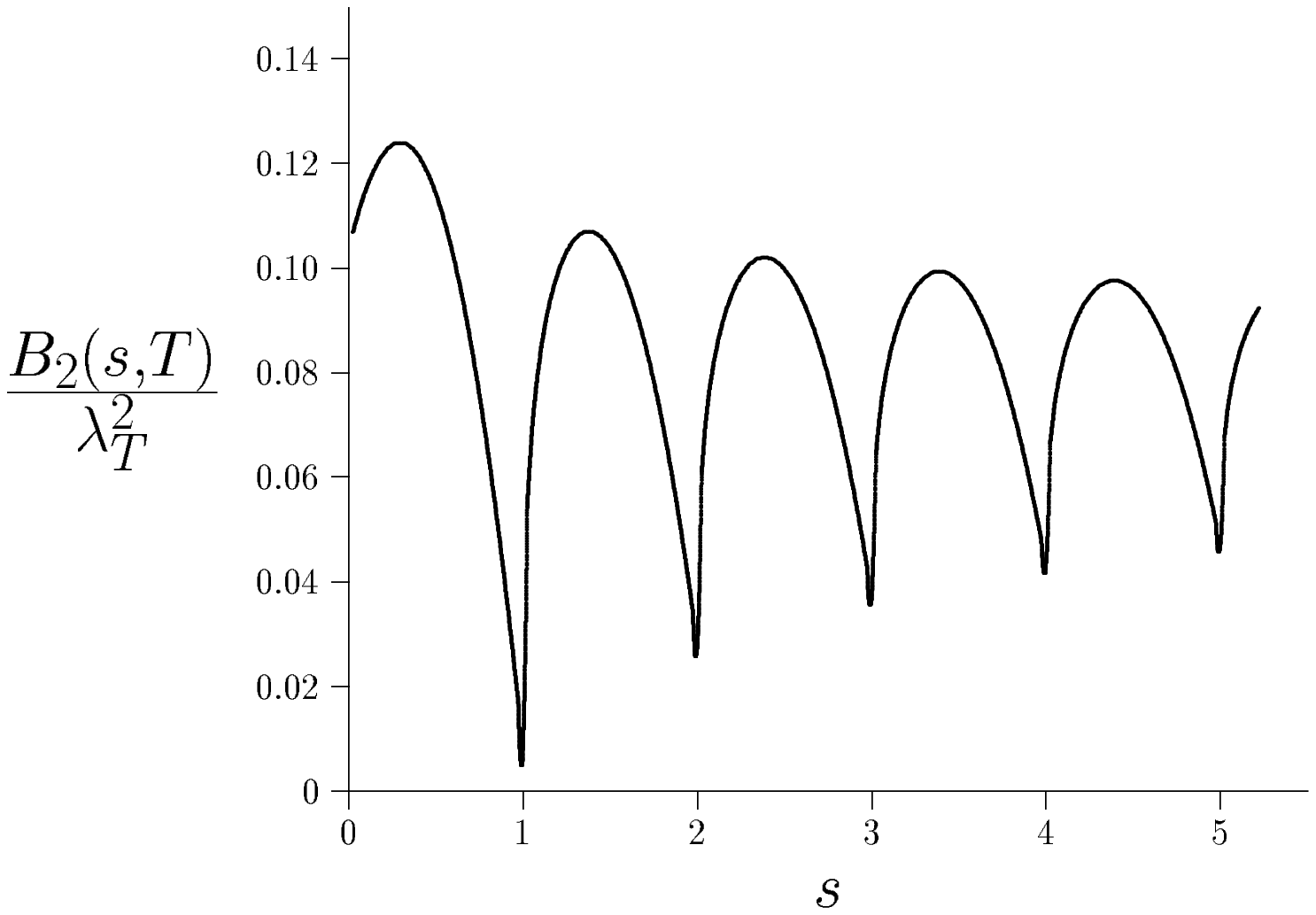,height=5cm}}
\bigskip
\caption{The second virial coefficient  as a function of the
induced spin $s$ in the large-$\kappa$ limit.}
\end{figure}

\end{document}